\documentclass[aps,prl,twocolumn]{revtex4-1}

\usepackage{amsmath, amsthm, amssymb}
\usepackage{graphicx,color,psfrag}

\usepackage{color}
\newcommand{\Lfun}[2]{{\cal L}^{(#1)}_{#2}}

\begin{document}

%\title{Incompleteness of B\"uttiker's Voltage Probe Concept}
%\title{Thermoelectric Corrections to Quantum Measurement}
\title{Thermoelectric Corrections to Quantum Voltage Measurement}

\author{Justin\ P.\ Bergfield}
\affiliation{Department of Chemistry, Northwestern University, Evanston, IL, 60208, USA}
\email{justin.bergfield@northwestern.edu}

\author{Charles\ A.\ Stafford}
\affiliation{Department of Physics, University of Arizona, 1118 East Fourth Street, Tucson, AZ 85721}

%\date{\today}
\begin{abstract}
A generalization of B\"uttiker's voltage probe concept for %nonequilibrium quantum systems at finite temperatures
nonzero temperatures
is an open third terminal of a quantum thermoelectric circuit.
An explicit analytic expression for the thermoelectric correction to an ideal quantum voltage measurement is derived, and interpreted in terms of local 
Peltier cooling/heating within the nonequilibrium system.  
The thermoelectric correction is found to be large (up to $\pm 24\%$ of the peak voltage) in a prototypical ballistic quantum conductor 
(graphene nanoribbon).
The effects of measurement non-ideality are also investigated.  
Our findings have important implications for precision local electrical measurements. % in nonequilibrium quantum systems, and imply that a precision voltage measurement {\em requires} a precision temperature measurement.
%It is shown that if the thermal coupling of the voltage probe to the ambient
%environment exceeds the thermal conductance quantum $\kappa_{0}$, then the predicted thermoelectric correction to the voltage is strongly suppressed.
\end{abstract}

\maketitle

%\section{Introduction}

Following the work of Engquist and Anderson \cite{Engquist81}, 
Markus B\"uttiker developed a paradigm \cite{Buttiker88,Buttiker89,Buttiker07} 
of quantum voltage measurement carried out by a probe consisting of a reservoir of non-interacting electrons 
%able to exchange particles with a system of interest through a local coupling.
coupled locally to a system of interest.
The probe exchanges electrons with the system until it reaches local electrical equilibrium with the system:
%B\"uttiker's voltage probe \cite{Buttiker88,Buttiker89}  (also \cite{Buttiker07}).  
%Condition that probe is in local electrical equilibrium with system:
\begin{equation}
I_p^{(0)}=0,
\label{eq:equil0}
\end{equation}
where $-eI_p^{(0)}$ is the mean electrical current flowing into the probe.
Once equilibrium is established, the chemical potential $\mu_p\equiv -eV_p$ of the probe constitutes a measurement of the local electrochemical
potential (and voltage $V_p$) within the nonequilibrium
quantum system \cite{Buttiker89}.
The condition (\ref{eq:equil0}) implies the probe has a large electrical input impedance, a necessary condition for a faithful voltage measurement.
Scanning potentiometers satisfying these conditions \cite{kalinin2007scanning} are now a mature technology, and
many experiments in mesoscopic electrical transport
utilize voltage probes % of a similar nature as %real, physical 
as circuit components \cite{Benoit86,Shepard92,Picciotto01,Gao05}. 
%A voltage probe is a real, physical component used in many mesoscopic experiments \cite{Benoit86,Shepard92,Picciotto01,Gao05}.

Although the average electric current into the probe is zero, electrons are constantly being emitted from the system into the probe, and replaced by
electrons from the probe reservoir whose quantum mechanical phase is uncorrelated with those emitted by the system.  In this way, such a voltage probe
serves as an {\em inelastic scatterer} \cite{Buttiker88}.  Indeed, much of the theoretical interest in B\"uttiker's model of a voltage probe is as 
a convenient way to introduce inelastic scattering in a quantum coherent conductor at the expense of introducing one additional electrical terminal.

B\"uttiker's early analysis \cite{Buttiker88,Buttiker89} was confined to systems at absolute zero temperature, where thermoelectric effects are absent.
Later, voltage probes at finite temperature were considered in the limit where the thermal coupling of the probe to the environment is large, so
that the probe remains at ambient temperature despite its coupling to the nonequilibrium quantum system \cite{Buttiker07}.  
This limit is consistent with the original analysis of 
Engquist and Anderson \cite{Engquist81}, which did not consider thermoelectric effects.

However, considered as a model of an inelastic scatterer, a voltage probe cannot be a steady-state source or sink of heat \cite{Jacquet09}.
This suggests that in generalizing the voltage probe concept \cite{Buttiker88,Buttiker89} to finite temperatures, the probe should be not only in local
electrical equilibrium, but also
in local thermal equilibrium with the system:
\begin{equation}
I_p^{(1)}=0,
\label{eq:equil1}
\end{equation}
where $I_p^{(1)}$ is the heat current flowing into the probe.
Condition (\ref{eq:equil1}) is required for a probe with a large thermal input impedance.

Further support for the additional condition (\ref{eq:equil1}) is provided by considering thermoelectric effects in the three-terminal circuit formed by the system with
source, drain, and probe.  Even if both source and drain electrodes are held at ambient temperature, an electrical bias between source and drain can drive Peltier 
cooling/heating within the system, resulting in hot and cold spots differing significantly from ambient temperature.  If the probe is not allowed to equilibrate 
thermally with the system under these conditions, a voltage will develop across the system-probe junction due to the Seebeck effect.  Then the probe voltage can 
no longer be %unambiguously 
interpreted as a measurement of the local electrochemical potential in the system.
We thus define an {\em ideal voltage measurement} as one satisfying both conditions (\ref{eq:equil0}) and (\ref{eq:equil1}).
A precision voltage measurement thus requires a simultaneous precision temperature measurement.

%Of course, it may be argued that as a practical matter, 
A significant challenge to achieving such an ideal voltage measurement is posed by
thermal coupling of the probe to the environment \cite{Majumdar99,Kim12,Bergfield2013demon,Bergfield14},
including to the system's lattice \cite{Buttiker07}, 
which may not be in local thermal equilibrium with the nonequilibrium electron system.  Furthermore, this coupling may be many times
as large as the probe's local thermal coupling to the system's electrons \cite{Kim12,Bergfield2013demon,Bergfield14}.
The probe's thermal coupling to anything other than the nonequilibrium electron system of interest %must be considered a non-ideality that 
leads to a deviation of
the probe's voltage from the ideal value associated with the local electrochemical potential of the system, and thus must be considered a non-ideality.
The probe's thermal coupling to the system's lattice can be minimized when it is operated in the tunneling regime \cite{Bergfield2013demon}, and continued
advances in scanning thermal microscopy (SThM) \cite{Majumdar99,Kim11,Yu11,Kim12,Fabian12} promise to further reduce the probe's thermal coupling to the environment.

\section{Linear thermoelectric response}

In the limit of small electric and thermal bias away from the equilibrium temperature $T_0$ and chemical potential $\mu_0$, 
the electric current $-eI^{(0)}_p$ and heat current $I^{(1)}_p$ flowing into the probe %electrode 
may be expressed as \cite{Bergfield2013demon}
\begin{align}
	I_p^{(\nu)} &=  \Lfun{\nu}{p1}\left(\mu_1-\mu_p\right) +\Lfun{\nu}{p2} \left(\mu_2-\mu_p\right) 
 \nonumber \\ & +\Lfun{\nu+1}{p1}\left(\frac{T_1-T_p}{T_0}\right)+\Lfun{\nu+1}{p2}\left(\frac{T_2-T_p}{T_0}\right) \nonumber \\ &
  \phantom{yoyoyo}+\delta_{\nu,1} \kappa_{p0}(T_0-T_p),
\label{eq:three_terminal_currents}
\end{align}
where 
$\Lfun{\nu}{\alpha\beta}$ are Onsager linear-response coefficients with electrode labels $\alpha$ and $\beta$, and $\kappa_{p0}= \Lfun{2}{p0}/T_0$ 
is the thermal conductance between the probe and the ambient environment %(which is assumed not to have any electrical sources or sinks) 
\cite{Bergfield2013demon}.  Eq.\ (\ref{eq:three_terminal_currents}) is a {\em completely general} linear-response formula, and applies to macroscopic systems, mesoscopic systems, nanostructures, etc., including electrons, phonons, and all other degrees of freedom, with arbitrary interactions between them.  

At sufficiently low temperatures or for sufficiently small systems, the electronic contribution to the coefficients $\Lfun{\nu}{\alpha\beta}$
may be calculated using elastic quantum transport theory \cite{Sivan86,Bergfield09b,Bergfield10}
\begin{equation}
\Lfun{\nu}{\alpha\beta}%\left(\mu,T\right) 
= \frac{1}{h} \int dE \; (E-\mu_0)^{\nu}\,T_{\alpha\beta}(E) \left(-\frac{\partial f_0}{\partial E}\right),
\label{eq:Lnu_elastic}  
\end{equation}
where $T_{\alpha\beta} (E)$ is the quantum mechanical transmission function \cite{Datta95} describing the probability to propagate from 
electrode $\beta$ to electrode $\alpha$, and %$f_0$ 
\begin{equation}
f_0(E) = \frac{1}{\exp\left(\frac{E-\mu_0}{k_B T_0}\right) + 1}
\end{equation}
is the equilibrium Fermi-Dirac distribution. % of the electrodes at chemical potential $\mu_0$ and temperature $T_0$.

\subsection{B\"uttiker's voltage probe}

In the limit as the system temperature approaches absolute zero, Eq.\ (\ref{eq:Lnu_elastic}) becomes
\begin{equation}
\lim_{T_0 \rightarrow 0} \Lfun{\nu}{\alpha\beta} = \frac{1}{h} T_{\alpha\beta}(\mu_0) \delta_{\nu,0}.
\label{eq:Lnu_0}
\end{equation} 
Then Eqs.\ (\ref{eq:equil0}) and (\ref{eq:three_terminal_currents}) may be solved to obtain 
B\"uttiker's result \cite{Buttiker88,Buttiker89} for the voltage measured by the probe 
\begin{equation}
\mu_p^{\rm B} \equiv \lim_{T_0\rightarrow 0} \mu_p = \frac{T_{p1}(\mu_0) \mu_1 + T_{p2}(\mu_0) \mu_2}{T_{p1}(\mu_0)+ T_{p2}(\mu_0)}.
\label{eq:mu_p_MB}
\end{equation}

\subsection{Engquist and Anderson's voltage probe}

The question remains, how to generalize B\"uttiker's result (\ref{eq:mu_p_MB}) to systems at non-zero temperatures.
%The voltage/temperature probes proposed by Engquist and Anderson \cite{Engquist81} 
Early on, Engquist and Anderson \cite{Engquist81} considered both voltage and temperature probes of quantum electron systems at finite temperature.
For the case of a voltage measurement, they assumed the entire system remains at ambient temperature $T_1=T_2=T_p=T_0$, so that 
Eqs.\ (\ref{eq:equil0}) and (\ref{eq:three_terminal_currents}) imply
\begin{equation}
	\mu_p^{\rm EA} = \frac{\Lfun{0}{p1} \mu_1 + \Lfun{0}{p2} \mu_2}{\Lfun{0}{p1} +\Lfun{0}{p2}}.
	\label{eq:EA_voltage}
\end{equation}
%where $\Lfun{\nu}{\alpha\beta}$ are Onsager linear-response coefficients with electrode labels $\alpha$ and $\beta$.  
However,
substituting Eq.\ (\ref{eq:EA_voltage}) for the probe's chemical potential into Eq.\ (\ref{eq:three_terminal_currents}) gives
\begin{equation}
	I_p^{(1)} = \frac{\Lfun{1}{p1}\Lfun{0}{p2} - \Lfun{1}{p2}\Lfun{0}{p1}}{\Lfun{0}{p1}+\Lfun{0}{p2}}\left(\mu_1 - \mu_2 \right),
\label{eq:Ip1_EA}
\end{equation}
which is generally non-zero at finite temperature.
This is a generic three-terminal thermoelectric effect occuring %at finite temperature 
whenever the probe coupling to the source and drain electrodes
(through the system) is unequal.
%which is generally non-zero at finite temperature.
Thus the voltage probe originally proposed by Engquist and Anderson is {\em not in thermal equilibrium} with the system.
In the absence of thermal equilibrium, the identification of $\mu_p^{\rm EA}$ with the local electrochemical 
potential of the system is problematic, since any temperature
differential between sample and probe will lead to a voltage differential through the Seebeck effect.
Moreover, the assumption that $T_p=T_0$ is inconsistent,
given that $I_p^{(1)}\neq 0$,
unless %$\kappa_{p0} \gg ?$
the thermal coupling of the probe to the environment is so large that the heat current flowing into the probe from the system can be neglected.

\begin{figure*}[htb]
	\centering
	\includegraphics[width=4in]{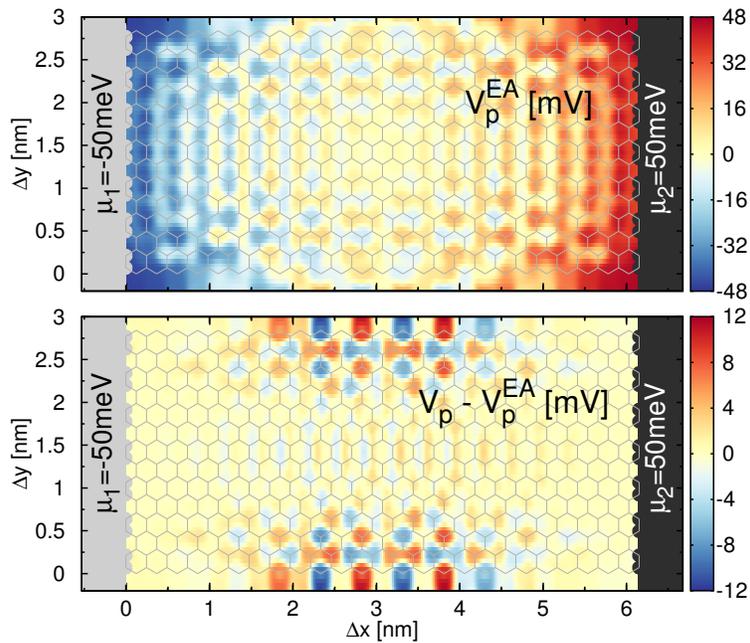}  
			\caption{The calculated response of a voltage probe scanned 3\AA\  above the plane of a zig-zag graphene nanoribbon. 
Top panel: The voltage distribution calculated using Engquist and Anderson's theory \cite{Engquist81}, cf.\ Eq.\ (\ref{eq:EA_voltage}). 
This theory neglects thermoelectric effects.  
The peak voltage for this system is 47.6mV. Bottom panel: The thermoelectric correction $\Delta V_p$ to the probe voltage, calculated
using Eqs.\ (\ref{eq:Delta_Vp}--\ref{eq:kappa_p}), reaching a maximum value for this system of 11.5mV.  
%The probe's voltage correction profile does not match that of the voltage itself and represents a significant source of error ($\sim$24\% of 
%the peak measured voltage, and $\sim$11.5\% of the applied voltage) in any {\em precision} voltage measurement.  
%This error is still significant even for non-ideal measurments (cf. Fig\ \ref{fig:CNR_voltage_temp_with_kappa_p0}.  
Calculations are performed at $\mu_0-\mu_{\rm Dirac}=-57.5\mbox{meV}$, 
$\mu_2-\mu_1=0.1\mbox{eV}$, $T_1=T_2=T_0=300\mbox{K}$, and $\kappa_{p0}=0$.}
	\label{fig:CNR_voltage_voltage_error}
\end{figure*}

\section{Ideal Voltage Measurement}

% Conditions of validity for the below equation are $T_1=T_2=T_0$
We define an {\em ideal voltage measurement} as one in which the probe is in {\em both electrical and thermal equilibrium} with the system.  
For an electrical bias $\Delta \mu=\mu_1-\mu_2$ applied between electrodes 1 and 2, both held at ambient temperature ($T_1=T_2=T_0$),
Eqs.\ (\ref{eq:equil0}--\ref{eq:three_terminal_currents})
%, (\ref{eq:equil1}), and (\ref{eq:three_terminal_currents}) 
can be solved for the probe voltage of such an ideal measurement, yielding
$\mu_p=\mu_p^{\rm EA} -e\Delta V_p$, where 
the
thermoelectric correction to the voltage is %inherent in an ideal %quantum voltage measurement
\begin{equation}
%\Delta V_p= \frac{S_{ps} \, I_p^{(1)}}{\kappa_{ps}+\kappa_{p0}}, %\frac{1}{1+\kappa_{p0}/\kappa_{ps}},
\Delta V_p= S_{ps}(T_p-T_0),
\label{eq:Delta_Vp}
\end{equation}
%where 
\begin{equation}
S_{ps}=-\frac{1}{eT_0} \frac{\Lfun{1}{p1}+\Lfun{1}{p2}}{\Lfun{0}{p1}+\Lfun{0}{p2}}
\label{eq:S_ps}
\end{equation}
is the thermopower of the probe-sample junction, and $T_p$ is the probe temperature satisfying
\begin{equation}
        T_p-T_0 = \frac{I_p^{(1)}}{\kappa_{ps}+\kappa_{p0}},
        \label{eq:Tp}
\end{equation}
where $I_p^{(1)}$ is given by Eq.\ (\ref{eq:Ip1_EA}),
\begin{equation}
	\kappa_{ps} = %\kappa_{p0} + 
\frac{1}{T_0} \left[\left(\Lfun{2}{p1}+\Lfun{2}{p2}\right) - \frac{\left(\Lfun{1}{p1}+\Lfun{1}{p2}\right)^2}{\left(\Lfun{0}{p1}+\Lfun{0}{p2}\right)}\right]
\label{eq:kappa_p}
\end{equation}
is the parallel thermal conductance from electrodes 1 and 2 %and the thermal environment 
into the probe, and $\kappa_{p0}$ is the thermal coupling of the probe to the environment at temperature $T_0$.

% Expand discussion?
%\JPB{Something like?}
%Brief discussion of relation to LDOS. This is three-terminal and the probe images the {\em partial} density of states.
%We don't just image the LDOS like an equilibrium system (i.e. a two-terminal setup with the probe and substrait only).

\section{Results}

In this section, we calculate the thermoelectric correction to the probe voltage for a prototypical ballistic quantum conductor, a graphene nanoribbon.  
However, we emphasize that the voltage error induced by thermoelectric effects is a generic phenomenon, and not material specific.  Figure 
\ref{fig:CNR_voltage_voltage_error} shows the computed voltage distribution %response and the thermoelectric correction to the voltage 
for a zig-zag graphene nanoribbon with an electrical bias of 0.1V between the source and
drain electrodes (at right and left in the figure), which are held at the ambient temperature of $T_0=300\mbox{K}$.  The equilibrium chemical 
potential of the nanoribbon  (determined by doping and/or a backgate)
was taken as $\mu_0-\mu_{\rm Dirac}$=-57.5meV.  %-57.51meV.  CAS: Too many sig-figs!
In our calculations, the $\pi$-system of the graphene nanoribbon is described using a tight-binding model which has been shown to accurately reproduce the 
low-energy physics of this system \cite{reich2002tight}.  The macroscopic electrodes are assumed to operate in the broad-band limit, where the 
electrode-nanoribbon coupling is independent of energy, with a per-orbital bonding strength of 2.5eV.  The voltage probe is modeled as an 
atomically-sharp Pt tip scanned at a fixed height of 3\AA\ above the plane of the C nuclei (tunneling regime).  The tunneling matrix elements between the 
probe atoms and the nanoribbon were determined using the methods outlined in Ref. \citenum{Chen93}.   The linear-response coefficients were calculated 
using Eq.\ (\ref{eq:Lnu_elastic}) following the methods of Refs.\ \cite{Bergfield2013demon,Bergfield14}.  Additional details of our computational methods may be found in the Supporting Information.

The top panel of Fig.\ \ref{fig:CNR_voltage_voltage_error} shows the Engquist-Anderson voltage $V_p^{\rm EA}\equiv -\mu_p^{\rm EA}/e$ 
computed from Eq.\ (\ref{eq:EA_voltage}),
%using $\mu_p^{\rm EA}\equiv -eV_p^{\rm EA}$.  
while the bottom panel %of Fig.\ \ref{fig:CNR_voltage_voltage_error} 
shows the thermoelectric correction $\Delta V_p$ to the probe voltage, 
computed from Eqs.\ (\ref{eq:Delta_Vp}--\ref{eq:kappa_p}).
%\JPB{This voltage resolution is feasable, and voltage measurements with atomic resolution are frequently achieved} 
For this case, which is representative of various geometries we have considered (See Supporting Information), the thermoelectric
correction to the measured voltage is $\pm 24\%$ of the maxiumum voltage and $\pm 11.5\%$ of the applied bias, highlighting the importance of 
thermoelectric effects on precision voltage measurements in quantum systems. As mentioned previously, this system is not unique and even larger corrections are expected for systems with larger thermoelectric responses.

\begin{figure}[htb]
	\centering
	\includegraphics[width=3.5in]{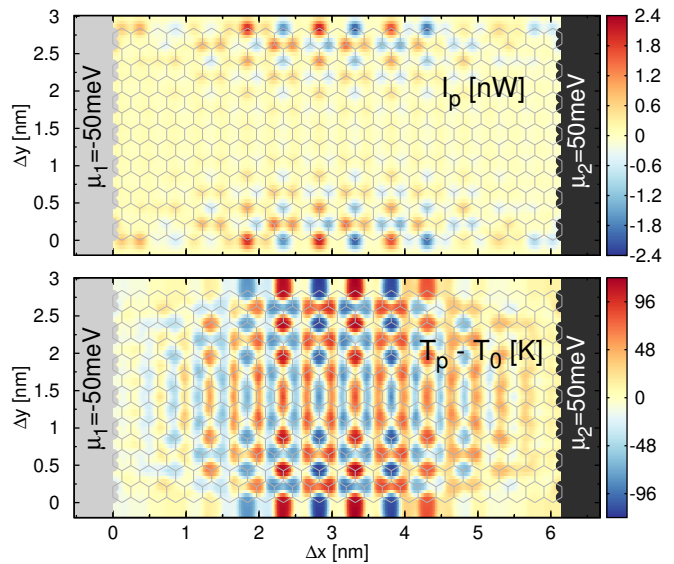}  
			\caption{Top panel: The heat current $I_p^{(1)}$ flowing into the probe when it is held at the ambient temperature $T_0=300\mbox{K}$,
calculated from Eq.\ (\ref{eq:Ip1_EA}).  
Bottom panel: The temperature $T_p$ of the probe when it is in both electrical and thermal equilibrium with the nonequilibrium electron system 
in the graphene nanoribbon, calculated from Eq.\ (\ref{eq:Tp}).  
%Although $T_p$ is assumed to be identical to $T_1=T_2=T_0$ in the Engquist-Anderson analysis \cite{Engquist81}, clearly it is not.  The origin of the large error in the probe's predicted temperature measurement (peak error is 119K) are thermoelectric effects, which often play an important role in nanostructures operating above absolute zero when the transport has a strong energy dependence.
}
	\label{fig:CNR_current_temp_diff}
\end{figure}

The cause of the substantial thermoelectric correction to the voltage %measurement 
is elucidated in Fig.\ \ref{fig:CNR_current_temp_diff}.
The top panel of Fig.\ \ref{fig:CNR_current_temp_diff} shows the heat current $I_p^{(1)}$ flowing into the probe when its temperature is
held fixed at $T_0$, calculated using Eq.\ (\ref{eq:Ip1_EA}).  The peak values of $I_p^{(1)}=\pm 2.3\,\mbox{nW}$ may not be large in an absolute sense, but
they correspond to a heat current density of $j_p^{(1)}=4.5\times 10^{10}\, \mbox{W}/\mbox{m}^2$ through the apex atom at the tip of the probe, 
some %\JPB{five} orders of magnitude greater than the intensity of the noonday sun in the Summer in Tucson! 
700 times the radiant energy flux at the surface of the sun!
Clearly, the assumption that such a probe, whose voltage is given
by Eq.\ (\ref{eq:EA_voltage}), is in local equilibrium with the system is questionable.

The bottom panel of Fig.\ \ref{fig:CNR_current_temp_diff}
shows the deviation of the temperature $T_p$ of an ideal thermoelectric probe from ambient temperature, calculated from Eq.\ (\ref{eq:Tp}).  
The ideal probe is in local thermal equilibrium with 
the system, and as such, its temperature maps out the hot and cold regions of the system \cite{Bergfield2013demon,Meair14,Bergfield14}.  
The lower panel of Fig.\ \ref{fig:CNR_current_temp_diff} shows
clear evidence of Peltier cooling/heating of up to $\pm 100\mbox{K}$ within the system induced by the external electrical bias of 0.1V.  
The large Peltier effect in this system may be related to giant thermoelectric effects predicted in 
related $\pi$-conjugated systems \cite{Bergfield09b,Bergfield10}, where quantum interference effects have been shown to strongly enhance thermoelectricity. 
However, similar phenomena should occur in other ballistic quantum conductors.
% Include some details of calculation

% Description of Error

%
%\begin{figure*}[tb]
%	\centering
%	%	\includegraphics[width=6in]{2014_0527_11x16_Mu_-0.05_current.eps}  %--> separate figure?
%	%	\includegraphics[width=6in]{1220_11x16_Mu_-0.1_voltage_error_dMu_.1eV.eps}
%	%	\includegraphics[width=6in]{1220_11x16_Mu_-0.1_probe_temp_dMu_.1eV.eps}
%		\caption{
%The calculated response of a thermoelectric probe scanned {\color{blue} TO DO: 3\AA\ ?} above the plane of an armchair graphene nanoribbon.  
%Top panel: The heat current flowing into the probe when it is held at the ambient temperature $T_0=300\mbox{K}$.
%Middle panel: The relative thermoelectric correction to the probe voltage 
%$(\mu_p-\mu^{\rm EA}_p)/(\mu_2-\mu_1)$.  Bottom panel: The temperature of the probe when it is in both electrical and thermal equilibrium with the 
%nonequilibrium electron system in the graphene nanoribbon.
%Although $T_p$ is assumed to be idential to $T_1=T_2=T_0$ in the Engquist-Anderson analysis \cite{Engquist81}, clearly it is not.  
%Calculations are performed with $\mu_0-\mu_{\rm Dirac}$=-0.1eV, 
%$\mu_2-\mu_1$=0.1eV, $T_0$=300K and $\kappa_{p0}$=0. {\color{blue} TO DO: gate voltage for top panel=-0.1V? Electrical not thermal bias?}
%}
%	\label{fig:CNR_voltage_temp_intrinsic}
%\end{figure*}

\subsection{Effect of thermal coupling of probe to environment}

\begin{figure}[htb]
	\centering
	\includegraphics[width=3.5in]{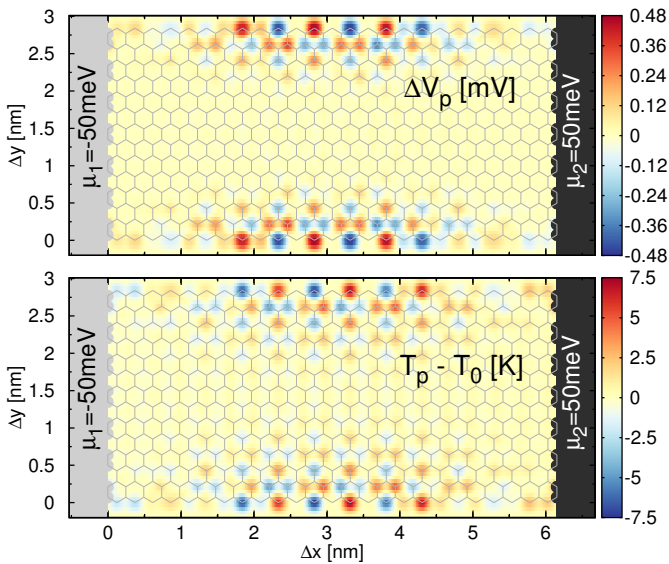}  
			\caption{The thermoelectric correction $\Delta V_p$ to the probe voltage (top panel) 
and the deviation of the probe temperature from ambient temperature (bottom panel) for the same system shown in Figs.\ 
\ref{fig:CNR_voltage_voltage_error}--\ref{fig:CNR_current_temp_diff}, 
but with a finite thermal coupling $\kappa_{p0}=\kappa_0$ of the probe to the environment, where  
$\kappa_0=(\pi^2/3)k_B^2 T_0/h$ (0.284nW/K at 300K) is the thermal conductance quantum.
Eqs.\ (\ref{eq:Delta_Vp}) and (\ref{eq:Tp}) indicate that 
the thermoelectric corrections for larger values of $\kappa_{p0}$ scale as $\kappa_{p0}^{-1}$.
%Although this value is consistent with current scanning voltage probe measurements \JPB{CITE}, these corrections decrease linearly with $\kappa_{p0}$.
}
	\label{fig:CNR_voltage_temp_with_kappa_p0}
\end{figure}

% Expand
Let us now consider the effects of measurement non-ideality.  %In particular, 
The greatest source of error in a scanning thermoelectric measurement is likely
to stem from the unavoidable %thermal 
coupling $\kappa_{p0}$ 
of the probe to the thermal background (typically, the ambient environment) \cite{Bergfield2013demon}.  Indeed, state-of-the-art SThM still operates in the
regime where the coupling of the probe to the thermal background is many times its thermal coupling to the system itself \cite{Kim12}.  While values of $\kappa_{p0}$ much less than the thermal conductance quantum $\kappa_0=(\pi^2/3)k_B^2 T_0/h$ (0.284nW/K at 300K) \cite{Rego98}
are possible in principle for probes whose thermal coupling to
the environment is predominantly radiative \cite{Bergfield2013demon}, current scanning probes \cite{Kim12} have $\kappa_{p0}> 100\kappa_0$.

%\JPB{Scanning probe potentiometry is a mature field and voltage measurements with $\mu V$ precision and atomic spatial resolution have been 
%Scanning tunneling potentiometry and scanning noise potentiometry have }

Figure \ref{fig:CNR_voltage_temp_with_kappa_p0} shows the thermoelectric correction to the voltage (upper panel) and the probe temperature 
(lower panel) for $\kappa_{p0}= \kappa_0$.  For this case, the thermal coupling of the probe to the environment %greatly 
exceeds its coupling to the system,
so that the probe temperature is closer to ambient, and the thermoelectric correction to the voltage is reduced.
The reduction of the thermoelectric corrections is described analytically by Eqs.\ (\ref{eq:Delta_Vp}) and (\ref{eq:Tp}).  Even for a thermal
coupling of $\kappa_{p0}=700\kappa_0$, typical of current state-of-the-art SThM \cite{Kim12}, the voltage error would still be of order 1$\mu$V, well
within the resolution of precision voltage measurements, which %\JPB{typically?} Not typically, many voltage measurements average over a macroscopic area
routinely obtain sub-\AA ngstrom spatial resolution \cite{kalinin2007scanning}.

%In this limit, the probe voltage approaches the value given by Engquist and
%Anderson's original formula \cite{Engquist81}, which neglects thermoelectric effects.
%However, the voltage error of tens of millivolts is still very large for precision measurements, which routinely obtain sub-\AA\ ngstrom spatial resolutions.

%\JPB{Voltage measurement and temperature measurement errors decrease linearly with increased $\kappa_{p0}$}

%\JPB{Another source of non-ideal measurement is reduced spatial resolution.  Although this is an important consideration, atomic resolution is currently achievable with existing voltage measurement (nanopotentiometry, }

%Scanning tunneling potentiometry can resolve the chemical potential to better than 100$\mu eV$ \cite{ramaswamy1998study}

%Scanning noise potentiometry measurements accuracy of $\sim 1\mu V$.\cite {moller1991scanning}  

\section{Conclusions}

%\JPB{Changed quite a bit}

An ideal voltage measurement in a nonequilibrium quantum system was defined in terms of a floating thermoelectric probe that reaches both electrical
and thermal equilibrium with a system via local (e.g., tunnel) coupling. This definition extends B\"uttiker's quantum voltage probe paradigm
\cite{Buttiker88,Buttiker89}
to systems at finite temperature, where thermoelectric effects are important. 

As an example, we
developed a realistic model of a scanning %electron 
potentiometer with atomic resolution and used it to investigate voltage measurement 
in a prototypical ballistic quantum conductor (a graphene nanoribbon) bonded to source and drain electrodes.  Under ideal measurement conditions, we predict large thermoelectric voltage corrections ($\sim$24\% of the probe's peak voltage signal) when the applied source-drain bias voltage is small.  We also derived expressions for the probe's voltage correction under non-ideal measurement conditions, finding that the voltage correction is reduced linearly as the probe-environment coupling is increased.  
In the graphene nanoribbon system considered here, voltage corrections on the order of several $\mu$V persist even with strong environmental coupling. 
%($\kappa_{p0} \lt 500\kappa_{p0}$).

In summary, we predict a large thermoelectric correction to voltage measurement in quantum coherent conductors.  The origin of this correction is 
local Peltier cooling/heating within the nonequilibrium quantum system, a generic three-terminal thermoelectric effect. %which are present in virtually all systems.  
This finding has important implications for precision local electrical measurements: it implies that {\em a precision voltage measurement requires
a simultaneous precision temperature measurement}.

%However, the predicted thermoelectric correction is strongly suppressed when
%the coupling of the probe to the thermal background exceeds the thermal conductance quantum.  

\begin{acknowledgments}
This paper is dedicated to the memory of Markus B\"uttiker, whose ideas continue to inspire our quest to understand nonequilibrium quantum systems.
%J.P.B.\ would like to thank Mark Ratner and Abe Nitzan for useful discussions.
C.A.S.\ was supported by the Department of Energy--Basic Energy Sciences grant no.\ DE-SC0006699.
\end{acknowledgments}

%\begin{suppinfo}
%Includes a discussion of the voltage correction for large probe-environment coupling and details of our computational methods.
%\end{suppinfo}

%merlin.mbs apsrev4-1.bst 2010-07-25 4.21a (PWD, AO, DPC) hacked
%Control: key (0)
%Control: author (8) initials jnrlst
%Control: editor formatted (1) identically to author
%Control: production of article title (-1) disabled
%Control: page (0) single
%Control: year (1) truncated
%Control: production of eprint (0) enabled
%

\end{document}